\documentclass[aps,pra,twocolumn,showpacs]{revtex4}
\usepackage{amsmath,amssymb,graphicx,bbold}
\usepackage{graphicx}  
\usepackage{dcolumn}   
\usepackage{bm}        
\usepackage{color}
\usepackage{braket}
\usepackage{cancel}
\usepackage{ulem}

\providecommand{\abs}[1]{\lvert #1 \rvert}
\providecommand{\norm}[1]{\lVert #1 \rVert}
\providecommand{\moy}[1]{\langle #1 \rangle}

\providecommand{\ket}[1]{\lvert #1 \rangle}
\providecommand{\braket}[2]{\langle #1 \rvert #2 \rangle}

\begin{document}
\title{Transfer of angular spectrum in parametric down-conversion with structured light}
%
\author{Antonio Z. Khoury$^1$, Paulo H. Souto Ribeiro$^2$, and Kaled Dechoum$^1$}

\affiliation{Instituto de F\'\i sica, Universidade Federal Fluminense,
24210-346 Niter\'oi, RJ, Brasil}
\affiliation{Departamento de F\'\i sica, Universidade Federal de Santa Catarina,
	88040-900 Florian\'opolis, SC, Brasil}

\date{\today}

\begin{abstract}
We develop the formal approach to the angular spectrum transfer in parametric down-conversion that allows pumping with a structured beam.
The scheme is based on an entangled photon source pumped 
by a laser beam structured with a vector vortex polarization profile. This creates a two-photon quantum state 
exhibiting polarization-dependent transverse correlations that can be accessed through coincidence measurements
on the spatially separated photons. The calculated correlations are shown 
to present a spin-orbit profile typical of vector beams, however, distributed on separate measurement regions. 
Our approach allows the design of the pump beam vector spatial structure and measurement strategies
for potential applications of these entangled states, such as in quantum communication.
\end{abstract}

\maketitle

\section{introduction}
\label{sec:intro}

Quantum technologies based on photonic devices require coherent control of different optical degrees of freedom. 
The interplay between polarization and transverse modes of a laser beam has been successfully exploited in a number 
of quantum and classical experiments. In the quantum domain we can quote different applications, such as 
simulations of quantum algorithms \cite{Oliveira2005,Souza2010,Pinheiro2013}, quantum random walks \cite{Forbes2013}, 
environment-induced entanglement \cite{Malena2009,Passos2018}, decoherence \cite{Obando2020}, non-Markovianity 
\cite{Passos2019}, quantum communications \cite{Milione2015} 
and quantum sensing \cite{Toppel2014,Berg-Johansen2015}. 
The structural non-separability between polarization and transverse modes has been approached from different points 
of view \cite{Simon2010,Holleczek2011,Qian2011,Leonardo2014,Ghose2014,Aiello2015,Forbes2015,Qian2016} and has been used for investigating 
important properties of entangled states \cite{Souza2007,BellUFF2010,Kagalwala2013}. Quantum inspired experiments in classical optics has led to 
exciting applications, such as the mode transfer between different degrees of freedom using the teleportation algorithm 
\cite{Hashemi2015,Silva2016,Guzman2016}. Tripartite non-separability has also been studied in classical optics using 
polarization, transverse and longitudinal modes\cite{Balthazar2016}. All these developments were considerably favored by the development of 
important tools for spin-orbit coupling in laser beams \cite{Nagali2009,Barreiro2010,Karimi2010,Cardano2012}. 
These developments made possible the implementation of alignment-free quantum cryptography with vector beams 
\cite{Souza2008,DAmbrosio2012,Guo2020}. The orbital angular momentum can be also combined with other degrees of freedom like optical path to generate hyper-entanglement between two quantum memories \cite{Zhang2016}.

In the quantum domain, the interplay between polarization and spatial coherence of entangled photon pairs was approached and has demonstrated quantum image control through polarization entanglement 
in spontaneous parametric down-conversion (SPDC)\cite{Santos2001,Juliana2003}. SPDC is  a reliable source of photon pairs 
entangled in different degrees of freedom \cite{Walborn2010}. 
The phase-matching conditions fulfilled 
by parametric interaction impose time-energy, space-momentum and polarization constraints that are at the 
origin of multiple quantum correlations characterizing entanglement between the generated photons. This 
multiple entanglement in different degrees of freedom is sometimes referred to as hyper-entanglement \cite{Barreiro05}. 

In this work we  present the theoretical description of photon pairs, simultaneously entangled in spin and orbital 
angular momentum, generated by the two-crystal-sandwich SPDC source \cite{Kwiat1999}  pumped by a vector vortex beam. The approach is an extension of the theory that describes the angular spectrum transfer in parametric down-conversion \cite{Monken1998}, for the case of structured beams.
The polarization dependent spatial correlations between the down-converted beams are calculated using the formal approach developed in Ref. \cite{Juliana2003}, where a setup with similar
characteristics was implemented. 
While the individual intensity distributions of the down-converted beams do not carry 
the pump spin-orbit properties, the quantum correlations between them exhibit the typical polarization dependent spatial distribution of a vector beam, as expected and experimentally observed in Ref. \cite{Samanta2017}. 
We explore and illustrate our approach by calculating the polarization-dependent transverse spatial correlations of the two-photon quantum state for a few interesting cases, including the correlations in the orbital angular momentum (OAM) basis.

\section{Experimental Scheme}
\label{sec:sketch}

\begin{figure}
	\includegraphics[scale=0.45]{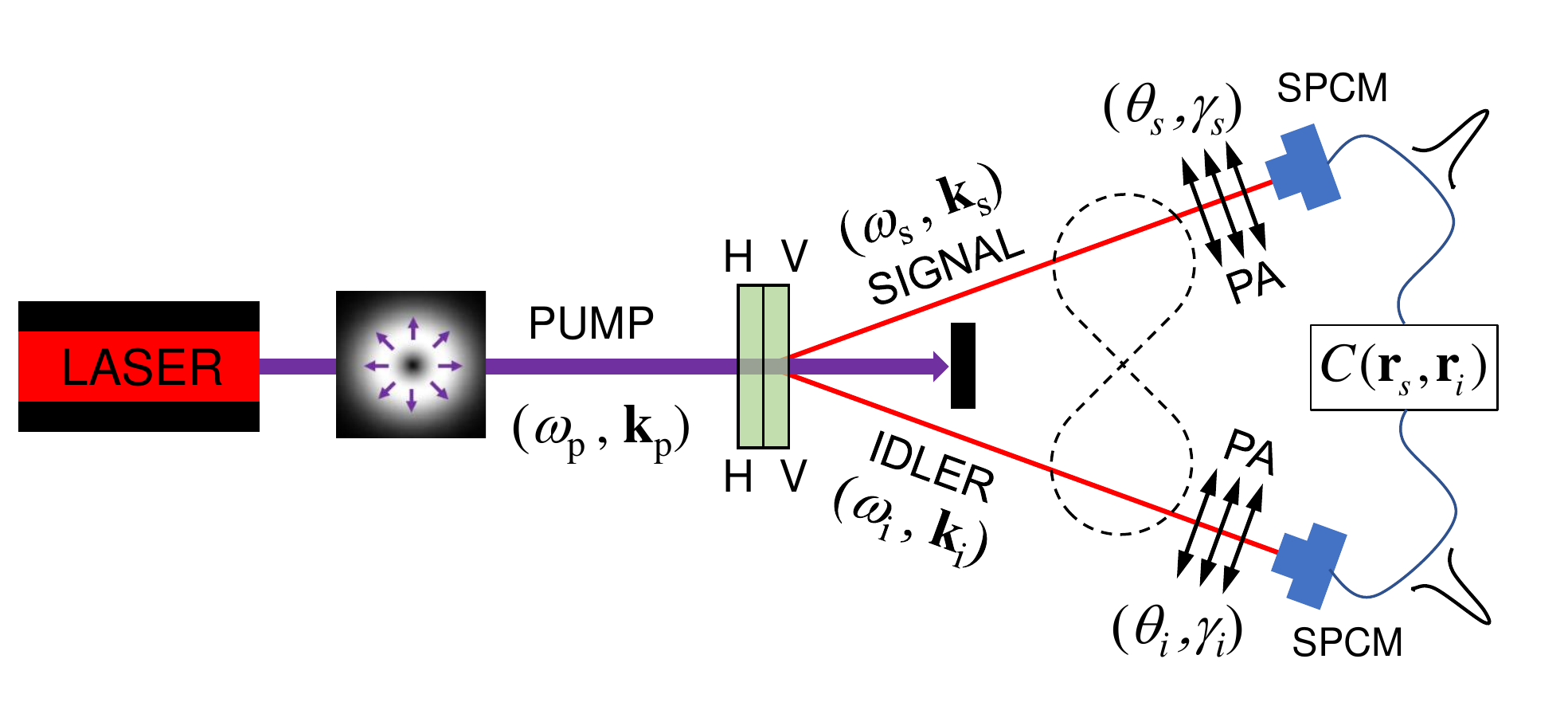}
	\caption{ Setup for the generation of two-photon vector vortex states. Two non-linear crystals cut for type-I phase-matching are  placed close together 
		with their optical axes rotated with respect to each other. A pump beam prepared in a vector-vortex mode can generate either horizontally 
		polarized photons in the first crystal or vertically polarized photons in the second. Before detection, a polarization analysis (PA) is performed in each detection arm. The polarization analyzers are set to  $(\theta_s/2 (HWP), \gamma_s (QWP))$ for the signal beam and $(\theta_i/2 (HWP) , \gamma_i (QWP))$ for the idler. After polarization analysis, 
		the down-converted photons hit the single-photon counting modules (SPCM) that can be scanned to register the polarization-dependent spatial correlations.}
	\label{fig:setup}
\end{figure}

Let us consider a frequently used source of polarization entangled photon pairs also known as {\it two-crystal-sandwich} SPDC source \cite{Kwiat1999}.
It is composed by two identical non-linear crystals placed close together with their optical axes rotated by $90^o$ relative to each other, 
as  shown in Fig. \ref{fig:setup}. 
A laser beam at frequency $\omega_p$ and wavector $\mathbf{k}_p$ is used to pump the crystals and generate photon pairs 
by spontaneous parametric down-conversion (SPDC). The down-converted signal ($s$) and idler ($i$) photons are generated with 
frequencies $\omega_s$ and $\omega_i\,$, and wave vectors $\mathbf{k}_s$ and $\mathbf{k}_i\,$, constrained by
$\omega_s + \omega_i = \omega_{p}$ and $\mathbf{k}_s + \mathbf{k}_i = \mathbf{k}_p\,$, that express energy and momentum 
conservation, respectively.

The pump beam is assumed to be prepared in a vector-vortex mode of the kind
\begin{equation}
\Psi(\mathbf{r}) = \frac{\psi_1(\mathbf{r}) \mathbf{\hat{e}}_H + \psi_2(\mathbf{r}) \mathbf{\hat{e}}_V}{\sqrt{2}}\;,
\label{vectormode}
\end{equation}
where $\Psi(\mathbf{r})$ is the classical field amplitude, $\mathbf{\hat{e}}_H$ and $\mathbf{\hat{e}}_V$ are the horizontal and vertical polarization unit vectors, 
respectively, and 
$\psi_1(\mathbf{r})$ and $\psi_2(\mathbf{r})$ are two orthonormal functions that are solutions of the paraxial wave equation \cite{yariv_1989}. 
Those can be either Hermite-Gaussian (HG) or Laguerre-Gaussian (LG) modes, for example. The LG modes are 
eigenstates of orbital angular momentum (OAM), that can be carried in multiples of $\hbar$ by each single photon.
The OAM carried by each photon is given by $l\hbar\,$, where $l\in\mathbb{Z}$ is the mode topological charge. 
The internal non-separability between the spin-orbit degrees of freedom can be  evidenced by measuring the spatial mode after polarization filtering. For example, if the vector-vortex mode given by \eqref{vectormode} passes through a 
polarization analyzer (a sequence of a quarter waveplate (QWP) and a half waveplate (HWP)followed by a horizontal polarizer, for example) 
characterized by angles $\gamma$ (QWP) and $\theta/2$ (HWP) with respect to the horizontal, the transmitted 
beam will exhibit a spatial function that is the linear combination
\begin{equation}
\psi_{\gamma\theta}(\mathbf{r}) = \cos\theta\,\psi_1(\mathbf{r}) + e^{i\gamma}\sin\theta\,\psi_2(\mathbf{r}) \;.
\label{rotatingvectormode}
\end{equation}
Therefore, a variable spatial profile is manifested after polarization projection. Interestingly, this feature can be 
transferred to the spatial quantum correlations between signal and idler photons generated by SPDC.

Under type-I phase-matching, the vertically polarized component of the pump beam generates a pair of horizontally polarized photons 
in the first crystal and transfers its accompanying transverse mode $\psi_1$ to the spatial quantum correlations of the 
down-converted photons \cite{Monken1998}. In the same way, the transverse mode $\psi_2$ is transferred to the spatial quantum correlations of the 
vertically polarized down-converted photons generated in the second crystal.  If the coherence length of the pump laser is larger than the length of the
two-crystal source, vertically and horizontally polarized signal-idler modes will add coherently and produce a two-photon vector vortex state.
A subtle and interesting effect takes place when the polarization information of the down-converted photons is erased by means of a variable polarization measurement. 
Two polarizers are used to set the measurement bases before the photocounts are acquired by 
single-photon counting modules (SPCM), which can be scanned to register the polarization-dependent spatial correlations. As we will show, these correlations 
exhibit typical features of vector beams, where polarization filtering is accompanied by a variable spatial profile. 
However, in our proposal this spin-orbit cross-talk is nonlocal. Note that these correlation images are generated by fixing either the signal or the idler position and scanning the other. Alternatively, the spatially dependent coincidence counts can be registered by currently available single-photon counting cameras. 

\section{Theoretical model}
\label{sec:theory}

Following the sketch of section \ref{sec:sketch}, we now develop a theoretical approach for the spin-orbit 
quantum correlations between signal and idler fields. Our strategy will be first to obtain the quantum state 
produced by the SPDC process and then use it to evaluate the spatial distribution of signal-idler intensity correlations.
Let us start by writing the positive and negative frequency parts of the electric field 
operator of pump, signal and idler beams as a superposition of plane waves with vertical and horizontal polarization 
\begin{eqnarray}
\mathbf{\hat{E}}^+_j(\mathbf{r},t) &=& 
\left(\hat{E}^+_{jH}(\mathbf{r})\,\mathbf{\hat{e}}_{H} + \hat{E}^+_{jV}(\mathbf{r})\,\mathbf{\hat{e}}_{V}\right)\,e^{-i\omega_{j}t}\;,
\nonumber\\
\mathbf{\hat{E}}^-_j(\mathbf{r},t) &=& \left[\mathbf{\hat{E}}^+_j(\mathbf{r},t)\right]^{\dagger}\;,
\\
\hat{E}^+_{j\mu}(\mathbf{r}) &=& i\,\mathcal{E}_{j} \int \hat{a}^j_{\mu}(\mathbf{k}_j)\,e^{i\mathbf{k}_j\cdot\mathbf{r}}\,d^3\mathbf{k}_j\;,
\nonumber
\end{eqnarray}
where $j=p,s,i\,$; $\hat{a}^j_{\mu}(\mathbf{k}_j)$ is the annihilation operator of photons with wave vector $\mathbf{k}_j$ 
and polarization $\mu=H,V\,$, and $\mathcal{E}_j$ 
is a constant resulting from the quantization process and having units of electric field.

The non-linear coupling between pump, signal and idler is described by the interaction Hamiltonian
\begin{eqnarray}
H_I(t) &=& \chi\,e^{-i\Delta\omega\,t}\,O_I + \chi\,e^{i\Delta\omega\,t}\,O_I^{\dagger}\;,
\label{hamiltonian}\\
O_I &=& \int_\mathcal{V} 
\left(\hat{E}^+_{pV}\,\hat{E}^-_{sH}\,\hat{E}^-_{iH} + \hat{E}^+_{pH}\,\hat{E}^-_{sV}\,\hat{E}^-_{iV}\right) d^3\mathbf{r}\;,
\nonumber
\end{eqnarray}
where $\chi$ is the non-linear susceptibility, $\mathcal{V}$ is the crystal volume and $\Delta\omega = \omega_{p}-\omega_{s}-\omega_{i}\,$.
The first term inside the integral describes the annihilation of 
a $V$ polarized photon of the pump and creation of $H$ polarized signal and idler photons in the first crystal. 
The second term describes the annihilation of an $H$ polarized photon of the pump and the creation of $V$ polarized signal and idler photons 
in the second crystal. 

\subsection{Two-photon spin-orbit quantum state}
\label{subsec:qstate}

Up to first-order in perturbation theory, the time evolution operator in the interaction picture is
\begin{eqnarray}
\!\!\!\!\!\!\!\!\!\!\!\!\!
&&U^{(1)}(\tau) = \mathbb{1}-\frac{i}{\hbar}\int_{0}^{\tau} H_I(t) \,dt
\label{U1}
\\
\!\!\!\!\!\!\!\!\!\!\!\!\!
&&= 
\mathbb{1}-\frac{i\chi}{\hbar}\,\frac{\sin\left(\Delta\omega\,\tau/2\right)}{\Delta\omega/2}\,
\left(e^{-\frac{i}{2}\Delta\omega\,\tau}\,O_I + e^{\frac{i}{2}\Delta\omega\,\tau}\,O_I^{\dagger}\right)\;,
\nonumber
\end{eqnarray}
where  $\tau$ is the interaction time, which is assumed to be the same for
all three fields, pump, signal, and idler, under phase-matching conditions. 
As we can see, the longer the interaction time, the tighter the energy conservation condition 
($\Delta\omega = 0$). At this point we can 
make the monochromatic approximation for the pump laser and assume the interaction time long 
enough to impose practically perfect energy conservation. In this case, the time evolution operator becomes
\begin{eqnarray}
U^{(1)}(\tau) = 
\mathbb{1}-\frac{i\chi\tau}{\hbar}\,
\left(O_I + O_I^{\dagger}\right)\;.
\label{U1-2}
\end{eqnarray}

\begin{figure}
	\includegraphics[scale=0.5]{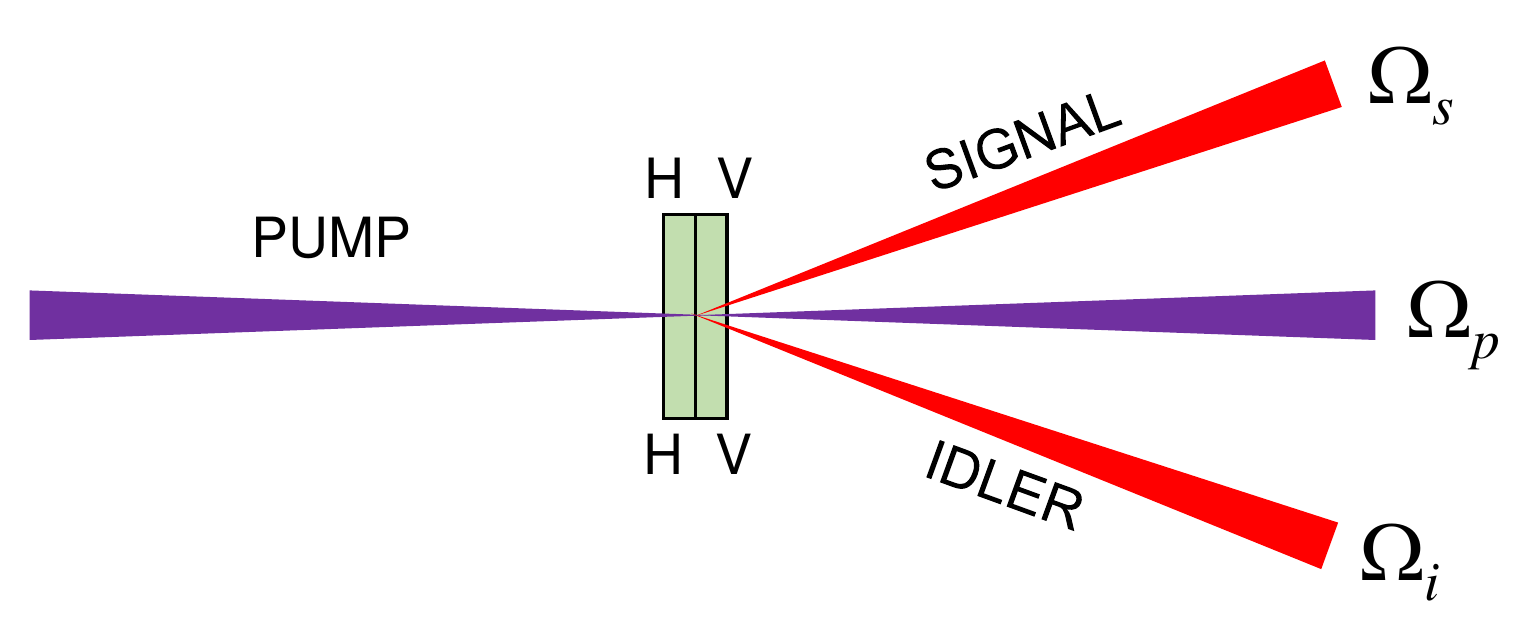}
	\caption{Representation of the transverse momentum domains for the pump, signal and idler beams.}
	\label{fig:conesdq}
\end{figure}

After passing through the crystals, the quantum state of the interacting beams is given by
\begin{eqnarray}
\ket{\Psi(\tau)} = U^{(1)}(\tau) \ket{\Psi(0)}\;,
\end{eqnarray}
where $\ket{\Psi(0)}=\ket{\psi_0}_p\ket{0}_s\ket{0}_i$ is the input state of pump, signal and idler fields. 
Since signal and idler are initially in the vacuum state, no contribution to the time evolution can appear from 
the $O_I^{\dagger}$ term because its action involves the annihilation of signal and idler photons.
The pump laser will be treated as a monochromatic beam, described by a multimode coherent state 
\begin{eqnarray}
\ket{\psi_0}_p &=& \prod_{\mathbf{k}_p}\ket{v_H(\mathbf{k}_p)}_H\ket{v_V(\mathbf{k}_p)}_V\;,
\end{eqnarray}
where
\begin{eqnarray}
\hat{a}_{\mu}(\mathbf{k})\ket{v_{\mu}(\mathbf{k})}_{\mu}\ &=& 
v_{\mu}(\mathbf{k})\ket{v_{\mu}(\mathbf{k})}_{\mu}\;,
\end{eqnarray}
and $v_{\mu}(\mathbf{k})$ is the coherent state amplitude associated with wave vector $\mathbf{k}$ and polarization $\mu\,$.
After parametric interaction, the quantum state of the pump, signal and idler modes is given by
\begin{eqnarray}
\ket{\Psi(\tau)} = \ket{\Psi(0)} -\frac{i\chi\tau}{\hbar}\,O_I \ket{\Psi(0)} \;.
\end{eqnarray}
Let us work out the interaction term:
\begin{eqnarray}
\!\!\!\!\!\!\!\!\!\!\!\!\!\!\!\!\!
&&O_I = 
-i\,\mathcal{E}_p\mathcal{E}_s\mathcal{E}_i \int d^3\mathbf{k}_p \int d^3\mathbf{k}_s \int d^3\mathbf{k}_i\, 
F(\mathbf{k}_p,\mathbf{k}_s,\mathbf{k}_i)
\nonumber \\
\!\!\!\!\!\!\!\!\!\!\!\!\!\!\!\!\!
&&\left[\hat{a}^p_V(\mathbf{k}_p)\hat{a}^{s\,\dagger}_H(\mathbf{k}_s)\hat{a}^{i\,\dagger}_H(\mathbf{k}_i) 
+
\hat{a}^p_H(\mathbf{k}_p)\hat{a}^{s\,\dagger}_V(\mathbf{k}_s)\hat{a}^{i\,\dagger}_V(\mathbf{k}_i)\right],
\end{eqnarray}
where we defined the phase-matching function as
\begin{eqnarray}
\!\!\!\!\!\!\!\!\!\!\!\!\!
F(\mathbf{k}_p,\mathbf{k}_s,\mathbf{k}_i)  \!&=&\!\!
\int_{\mathcal{V}} e^{i\left(\mathbf{k}_p - \mathbf{k}_s - \mathbf{k}_i\right)\cdot\mathbf{r}} \,d^3\mathbf{r}
\nonumber\\
\!\!\!\!\!\!\!\!\!\!\!\!\!
&\approx& \!\!\!
L_z\!\!
\prod_{l=x,y}\frac{2\sin\left[\left(k_{pl}-k_{sl}-k_{il}\right)L_l/2\right]}{k_{pl}-k_{sl}-k_{il}} ,
\end{eqnarray}
with $L_l$ being the crystal width along the $l$ direction. Note that we have assumed a longitudinally thin crystal 
satisfying $(k_{pz}-k_{sz}-k_{iz})L_z\ll 1\,$.
This results in 
\begin{eqnarray}
\!\!\!\!\!\!\!\!\!\!\!\!\!
&&O_I \ket{\Psi(0)} = 
\label{OIPsi}\\
\!\!\!\!\!\!\!\!\!\!\!\!\!
&&-i\,\mathcal{E}_p\mathcal{E}_s\mathcal{E}_i \int d^3\mathbf{k}_p \int d^3\mathbf{k}_s \int d^3\mathbf{k}_i\, 
F(\mathbf{k}_p,\mathbf{k}_s,\mathbf{k}_i)\,\ket{\psi_0}_p
\nonumber\\
\!\!\!\!\!\!\!\!\!\!\!\!\!
&&\otimes\left[\frac{}{}v^p_V(\mathbf{k}_p)\ket{1_{\mathbf{k}_s,H}}\ket{1_{\mathbf{k}_i,H}} 
+
v^p_H(\mathbf{k}_p)\ket{1_{\mathbf{k}_s,V}}\ket{1_{\mathbf{k}_i,V}}\right],
\nonumber
\end{eqnarray}
where $\ket{1_{\mathbf{k},\mu}}$ is a single-photon Fock state with wave vector $\mathbf{k}$ and polarization $\mu\,$. 
 We assume that the pump beam comes from a collimated and monochromatic laser propagating along the $z$ direction, so that we can 
approximate
\begin{eqnarray}
v^p_{\mu}(\mathbf{k}_p) \approx v^p_{\mu}(\mathbf{q}_p) \,\delta (k_{pz} -\, k_0)\,,
\end{eqnarray}
where $\mathbf{q}_p$ is the transverse and $k_0$ is the longitudinal wave vector component. Moreover, signal and idler photons 
are detected at small solid angles along specific directions compatible with the phase-matching condition. 
This geometric configuration, together with interference filters placed before the detectors, fix the 
selected wavelengths of signal and idler. This also 
restricts their detected wave vectors to a small neighborhood around their respective solid angles $\Omega_j$ 
($j=s,i$). It will be useful to decompose the wavectors into longitudinal ($\mathbf{k}^{\parallel}_{j}$) and transverse 
($\mathbf{q}_{j}$) parts 
$\mathbf{k}_{j} = \mathbf{k}^{\parallel}_{j} + \mathbf{q}_j\,$, and apply the paraxial approximation 
$|\mathbf{q}_{j}| \ll |\mathbf{k}^{\parallel}_{j}|\,$.
In this case, 
\begin{eqnarray}
k^{\parallel}_{j} = \sqrt{k_j^2 - q_j^2} 
\approx k_j - \frac{q_l^2}{2k_j}\,.
\label{paraxial}
\end{eqnarray}
In the paraxial regime and with fixed wavelengths for pump, signal and idler, the longitudinal wave vector 
components are fixed and the relevant plane wave modes can be labeled by the transverse wave vector 
$\mathbf{q}\,$. The amplitude distribution $v^p_{\mu}(\mathbf{q})$ represents the angular spectrum 
carried by the pump polarization mode $\mu\,$. The triple integrals in Eq. \eqref{OIPsi} are reduced to 
double integrals over a small domain $\Omega_j$ around the main longitudinal component, as 
illustrated in Fig. \ref{fig:conesdq}.
Under these assumptions, the two-photon quantum state can be written in terms of the transverse wavector 
distributions,
\begin{eqnarray}
\!\!\!\!\!\!\!\!\!
&&O_I \ket{\Psi(0)} = 
\\
\!\!\!\!\!\!\!\!\!
&&-i\,\mathcal{E}_p\mathcal{E}_s\mathcal{E}_i 
\int_{\Omega_p}\!\!\! d^2\mathbf{q}_p \int_{\Omega_s}\!\!\! d^2\mathbf{q}_s \int_{\Omega_i}\!\!\! d^2\mathbf{q}_i\, 
F(\mathbf{k}_p,\mathbf{k}_s,\mathbf{k}_i) \ket{\psi_0}_p
\nonumber \\
\!\!\!\!\!\!\!\!\!
&&\otimes\left[\frac{}{}v^p_V(\mathbf{q}_p)\ket{1_{\mathbf{q}_s,H}}\ket{1_{\mathbf{q}_i,H}} 
+
v^p_H(\mathbf{q}_p)\ket{1_{\mathbf{q}_s,V}}\ket{1_{\mathbf{q}_i,V}}\right]\,.
\nonumber
\end{eqnarray}
In most experiments for producing spatial quantum correlations in SPDC, one deals with transversely 
wide [$(k_{pl}-k_{sl}-k_{il})L_l\gg 1$ for $l=x,y$] and longitudinally thin [$(k_{pz}-k_{sz}-k_{iz})L_z\ll 1$] 
crystals. In this case, a tight phase-matching condition is imposed on the transverse components of the wave vectors 
and a loose condition applies to the longitudinal component. This approximation can be expressed as 
\begin{eqnarray}
F\left(\mathbf{k}_p,\mathbf{k}_s,\mathbf{k}_i\right) 
\approx L_z\, \delta\left(\mathbf{q}_p - \mathbf{q}_s - \mathbf{q}_i\right)\;.
\label{Fdelta}
\end{eqnarray}
In this case, the integration over $\mathbf{q}_p$ can be readily performed and 
the final expression for the quantum state produced by the SPDC process is
\begin{eqnarray}
\ket{\Psi(\tau)} = \ket{\psi_0}_p \otimes \left(\ket{0}_s\ket{0}_i + \ket{\Phi}_{si}\right)\;,
\end{eqnarray}
where
\begin{eqnarray}
\!\!\!\!\!\!\!\!\!\!\!\!
\ket{\Phi}_{si} \!\!&=&\!\! \kappa \int d^2\mathbf{q}_s \int d^2\mathbf{q}_i\, 
v^p_V(\mathbf{q}_s + \mathbf{q}_i)\ket{1_{\mathbf{q}_s,H}}\ket{1_{\mathbf{q}_i,H}} 
\nonumber\\
\!\!&+&\!\! 
\,\kappa \int d^2\mathbf{q}_s \int d^2\mathbf{q}_i\, 
v^p_H(\mathbf{q}_s + \mathbf{q}_i)\ket{1_{\mathbf{q}_s,V}}\ket{1_{\mathbf{q}_i,V}} \!,
\label{phitau}
\end{eqnarray}
and $\kappa \equiv -(\chi/\hbar)\tau L_z \mathcal{E}_p\mathcal{E}_s\mathcal{E}_i\,$.
The expression in \eqref{phitau} encompasses the transverse momentum constraint between signal and idler 
that gives rise to spatial quantum correlations. 

\subsection{Spatial quantum correlations}
\label{subsec:corr}

In the paraxial regime, we can adopt a scalar diffraction theory to study the propagation of the interacting 
beams after leaving the crystals. The relevant plane wave modes are labeled by the transverse wave vector 
$\mathbf{q}$ and the pump amplitude distribution $v^p_{\mu}(\mathbf{q})$ represents the pump angular spectrum 
carried by the polarization mode $\mu\,$. At the crystals' center ($z=0$), the spatial distribution of the pump 
beam in each polarization component is given by the following Fourier decomposition,
\begin{eqnarray}
W_{\mu}(\boldsymbol{\rho},0) = \int v^p_{\mu}(\mathbf{q}) \, e^{i\mathbf{q}\cdot\boldsymbol{\rho}} \, d^2\mathbf{q}\,,
\label{WFourier}
\end{eqnarray}
and the propagated field distribution is given by the Fresnel integral
\begin{eqnarray}
W_{\mu}(\boldsymbol{\rho},z) = e^{ikz}\int v^p_{\mu}(\mathbf{q}) \, 
e^{i\left(\mathbf{q}\cdot\boldsymbol{\rho}-\frac{q^2}{2k}z\right)} \, d^2\mathbf{q}\,.
\end{eqnarray}

The multimode coherent state $\ket{\psi_0}_p$ carries the spatial properties of the pump beam 
in the Fourier domain through the angular spectra $v_H(\mathbf{q})$ and $v_V(\mathbf{q})\,$, 
independently imprinted in each pump polarization component. This will be crucial for the 
polarization-dependent spatial correlations between signal and idler photons. 

The longitudinal positions of signal and idler detectors
are fixed and the spatial quantum correlations are measured as a function of 
their transverse position $\boldsymbol{\rho}_j$ ($j=s,i$). 
We recall that a polarization analyzer is placed before each detector.
If we assume that these analyzers are set at angles $\gamma_j$ and $\theta_j/2\,$, then the electric field 
operator in each detector is
\begin{eqnarray}
\hat{E}_j^{\,\prime}(\boldsymbol{\rho}_j) = \cos\theta_j\,\hat{E}_{jH}(\boldsymbol{\rho}_j) + \sin\theta_j\,e^{i\gamma_j}\,\hat{E}_{jV}(\boldsymbol{\rho}_j)\,.
\end{eqnarray}
The intensity distribution in each detection arm is
\begin{eqnarray}
I (\boldsymbol{\rho}_{j}) &=& 
\moy{ 
\hat{E}^{\,\prime -}_{j}(\boldsymbol{\rho}_j) \hat{E}^{\,\prime +}_{j}(\boldsymbol{\rho}_j)
}
\nonumber\\
&=& \norm{\hat{E}^{\,\prime +}_{j}(\boldsymbol{\rho}_j)\ket{\Psi(\tau)}}^2
\,,
\end{eqnarray}
and the intensity correlations between the two detection arms are
\begin{eqnarray}
C (\boldsymbol{\rho}_{s},\boldsymbol{\rho}_{i}) &=& 
\moy{ 
\hat{E}^{\,\prime -}_{s}(\boldsymbol{\rho}_s) \hat{E}^{\,\prime -}_{i}(\boldsymbol{\rho}_i) 
\hat{E}^{\,\prime +}_{i}(\boldsymbol{\rho}_i) \hat{E}^{\,\prime +}_{s}(\boldsymbol{\rho}_s)
}
\nonumber\\
&=& \norm{\hat{E}^{\,\prime +}_{i}(\boldsymbol{\rho}_i) \hat{E}^{\,\prime +}_{s}(\boldsymbol{\rho}_s)\ket{\Psi(\tau)}}^2\,.
\end{eqnarray}
In the monochromatic and paraxial approximations, the electric field operators can be written as
\begin{eqnarray}
\hat{E}^+_{j\mu}(\boldsymbol{\rho}) &=& i\,\mathcal{E}_{j} \int \hat{a}^j_{\mu}(\mathbf{q}_j)\,
e^{i\left[\mathbf{q}_j\cdot\boldsymbol{\rho} + \varphi(\mathbf{q}_j)\right]}
\,d^2\mathbf{q}_j\;,
\nonumber\\
\varphi(\mathbf{q}_j) &=& \sqrt{k_j^2 - q_j^2\,}\,\,z_j\approx k_j\,z_j - \frac{q_j^2z_j}{2k_j}\;,
\end{eqnarray}
where $z_j$ is the longitudinal distance between the crystals' center and detector $j$\,.

Note that no contribution comes from the vacuum component in $\ket{\Psi(\tau)}\,$, so that we only need 
to care about contributions coming from $\ket{\Phi}_{si}\,$.
The calculation of the intensity distributions and correlations will be significantly simplified by the 
definition of the following vectors
\begin{eqnarray}
\ket{\alpha_{j\mu}} &=& \hat{E}^+_{j\mu}(\boldsymbol{\rho}_j)\ket{\Phi}_{si}\;,
\nonumber\\
\ket{\beta_{\mu}} &=& \hat{E}^+_{i\mu}(\boldsymbol{\rho}_i)\hat{E}^+_{s\mu}(\boldsymbol{\rho}_s)\ket{\Phi}_{si}\;.
\label{betamu}
\end{eqnarray}
We can easily workout these auxiliary vectors using
\begin{eqnarray}
\hat{a}_{\mu}(\mathbf{q}^{\prime})\ket{n_{\mathbf{q},\nu}} = \delta_{\mu\nu}\delta(\mathbf{q}-\mathbf{q}^{\prime})
\sqrt{n_{\mathbf{q},\nu}}\ket{n_{\mathbf{q},\nu}-1}\;.
\end{eqnarray}
As detailed in Appendix \ref{individual}, the result for the individual intensities of 
signal and idler is 
\begin{equation}
I_j = \cos^2\theta_j \mathcal{I}_{H} + \sin^2\theta_j \mathcal{I}_{V}\,,
\label{eq:individual}
\end{equation}
where $j=s,i$ and 
\begin{equation}
\mathcal{I}_{H(V)} = \kappa^2\mathcal{E}_{s}^2 \, \int d^2\boldsymbol{\rho}
\abs{W_{V(H)}\left(\boldsymbol{\rho},z\right)}^2 \,.
\end{equation}
Note that the pump's spatial properties are washed out in the individual intensities.
In contrast, the coincidence count distribution carries the spatial profile of the 
pump beam, distributed in the joint coordinates of signal and idler. The result 
derived in Appendix \ref{coincidence} is 
\begin{eqnarray}
\!\!\!\!
C(\boldsymbol{\rho}_s,\boldsymbol{\rho}_i) &=& K^2 
\left|\frac{}{}\!\cos\theta_s\cos\theta_i\, W_{V}(\boldsymbol{\rho}_+,z) \right.
\nonumber\\
\!\!\!\!
&+& \left. \sin\theta_s\sin\theta_i\,e^{i(\gamma_s + \gamma_i)}\, W_{H}(\boldsymbol{\rho}_+,z)\right|^2 ,
\label{cwHV}
\end{eqnarray}
where we assume degenerate SPDC ($k_s = k_i = k_p/2$) and equidistant signal and idler detectors ($z_s = z_i =z$). 
In this case, we have
\begin{eqnarray}
\boldsymbol{\rho}_+ = \frac{\boldsymbol{\rho}_s + \boldsymbol{\rho}_i}{2}\,.
\end{eqnarray}
Note that Eq. \eqref{cwHV} shows simultaneous dependence on the joint coordinates of signal and idler detectors and 
on the joint orientations of their respective polarization analyzers.

\section{Nonlocal Vector Vortex Beam }
\label{sec:nonlocalvb}

We can now investigate the polarization dependent spatial correlations when the pump 
beam is prepared in a vector mode of the kind expressed in Eq.\eqref{vectormode}. For example, 
let the pump beam be prepared in a superposition of the Hermite-Gaussian mode $(0,1)$ with horizontal 
polarization and mode $(1,0)$ with vertical polarization so that
\begin{eqnarray}
&&W_{V}(\boldsymbol{\rho})=\frac{\psi_{10}(\boldsymbol{\rho})}{\sqrt{2}} = \frac{x}{\sqrt{\pi}\,w^2}\, e^{-(x^2+y^2)/w^2}\,,
\nonumber\\
&&W_{H}(\boldsymbol{\rho})=\frac{\psi_{01}(\boldsymbol{\rho})}{\sqrt{2}} = \frac{y}{\sqrt{\pi}\,w^2}\, e^{-(x^2+y^2)/w^2}\,.
\label{HGmodes}
\end{eqnarray}
The longitudinal dependence has been made implicit in the variation of the mode width 
$w(z) = w_0\sqrt{1+(z/z_R)^2}\,$, where $w_0$ is the mode waist and $z_R = \pi w_0^2/\lambda$ is the 
Rayleigh distance.

\begin{figure}
	\includegraphics[scale=0.5]{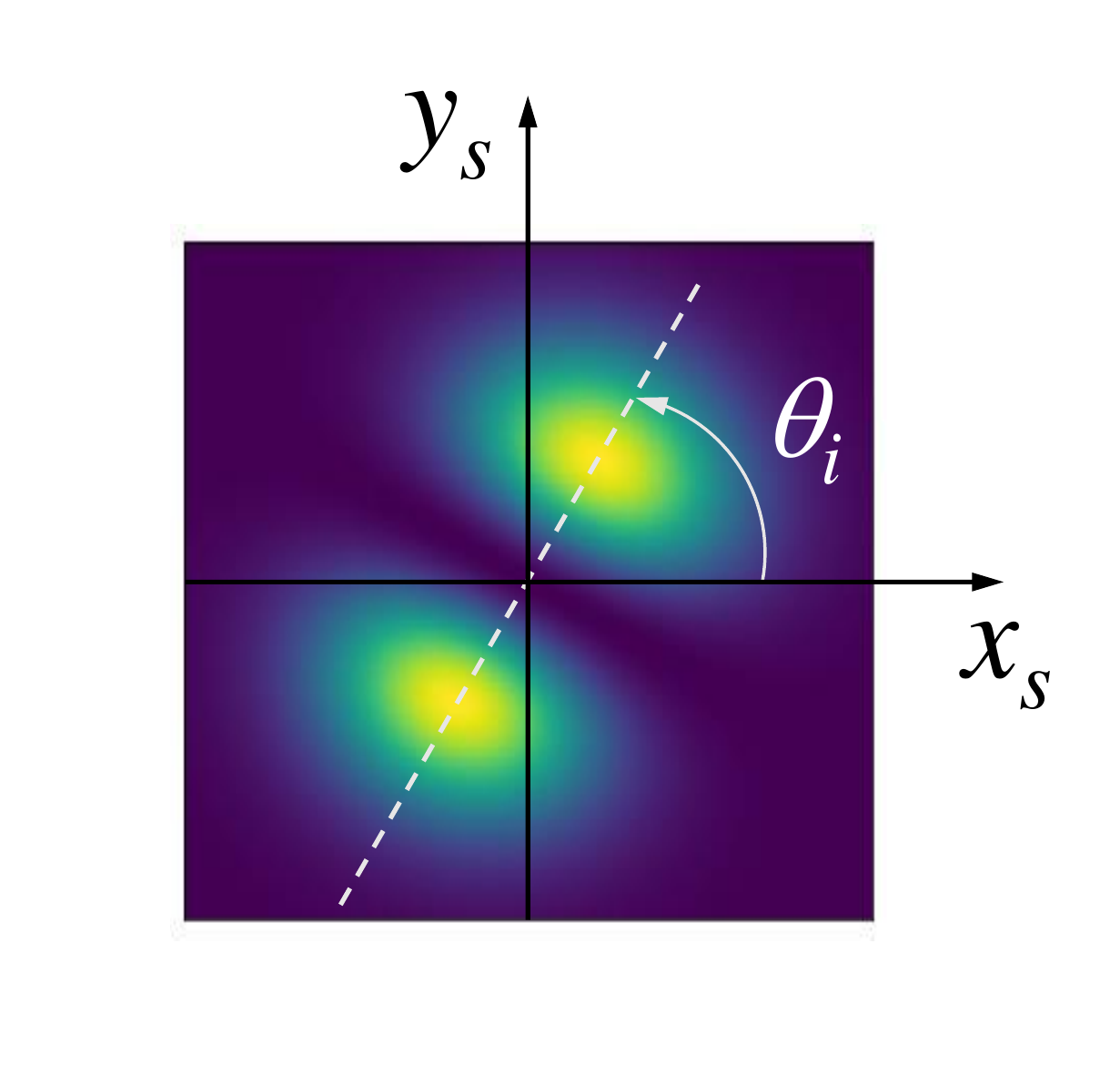}
	\caption{Hermite-Gaussian coincidence pattern as a function of the signal coordinates with the idler 
		detector fixed at the origin $x_i=y_i=0$ and the signal polarization analyzer fixed at 
		$\gamma_s = 0$ and $\theta_s = \pi/4\,$.
		The HG pattern follows the rotation of the idler polarization.}
	\label{fig:HGrotation}
\end{figure}

Then, the expression given in \eqref{cwHV} brings us to
\begin{eqnarray}
\!\!\!\!\!\!
&&C(\boldsymbol{\rho}_s,\boldsymbol{\rho}_i) =
\frac{K^2}{\pi\,w^4}\, e^{-\frac{2(x_s+x_i)^2}{w^2}} e^{-\frac{2(y_s+y_i)^2}{w^2}} \times
\label{CHG}\\
\!\!\!\!\!\!
&&
\left|\cos\theta_s \cos\theta_i\, (x_s+x_i) + \sin\theta_s\,\sin\theta_i\, e^{i(\gamma_s+\gamma_i)} (y_s+y_i)\right|^2 .
\nonumber
\end{eqnarray}
This expression shows simultaneous nonlocal behavior on position and polarization settings. The spatial distribution 
of the coincidence counts depends on the joint orientations of the detection polarizers and on the joint transverse 
positions of the detectors. The physical consequence of this double nonlocal behavior can be revealed by a simple 
measurement strategy. Let us set the signal polarization analyzer at $\gamma_s=0$ and $\theta_s = \pi/4\,$, 
so that the polarization information of the 
signal photons is erased, and the position of the idler detector kept at its origin $x_i=y_i=0\,$. In this case, 
the resulting coincidence pattern becomes
\begin{equation}
C(\boldsymbol{\rho}_s,\mathbf{0}) =
\frac{K^2}{2\pi\,w^4}\, e^{-\frac{2(x_s^2+y_s^2)}{w^2}} \!
\left(\cos\theta_i\, x_s + \sin\theta_i\, e^{i\gamma_i} y_s\right)^2 .
\label{cnonlocal}
\end{equation}
We can see that the resulting coincidence pattern corresponds to the intensity distribution of a first-order 
Hermite-Gaussian mode function of the signal position, transformed according to the parameters of the idler polarization 
analyzer. In Fig. \ref{fig:HGrotation} we plot this coincidence 
pattern as a function  of the signal coordinates $\boldsymbol{\rho}_s = (x,y)$, indicating that it follows
the rotation of the idler polarization analyzer.
This situation is similar to the 
one exhibited in Eq.\eqref{vectormode}, where the spatial profile after transmission of a single vector beam through 
a polarizer depends on the transmission angle. However, here we have the orientation of the spatial pattern determined 
by the angle of a remote polarizer. 
This effect can be useful for remote alignment of quantum cryptography 
stations or as a gyroscope. For example, in Refs. \cite{Souza2008,DAmbrosio2012} it was demonstrated that the internal non-separability between the spin and orbital degrees of freedom can be used to implement alignment-free quantum cryptography, thanks to the rotational invariance of spin-orbit modes. However, this method has never been considered in connection with quantum cryptography protocols employing non-local polarization correlations \cite{Ekert1991}. The non-local spin-orbit correlations described here can be useful in this context.

\section{Pumping with Laguerre-Gaussian beams}

\begin{figure}
	\includegraphics[scale=0.265]{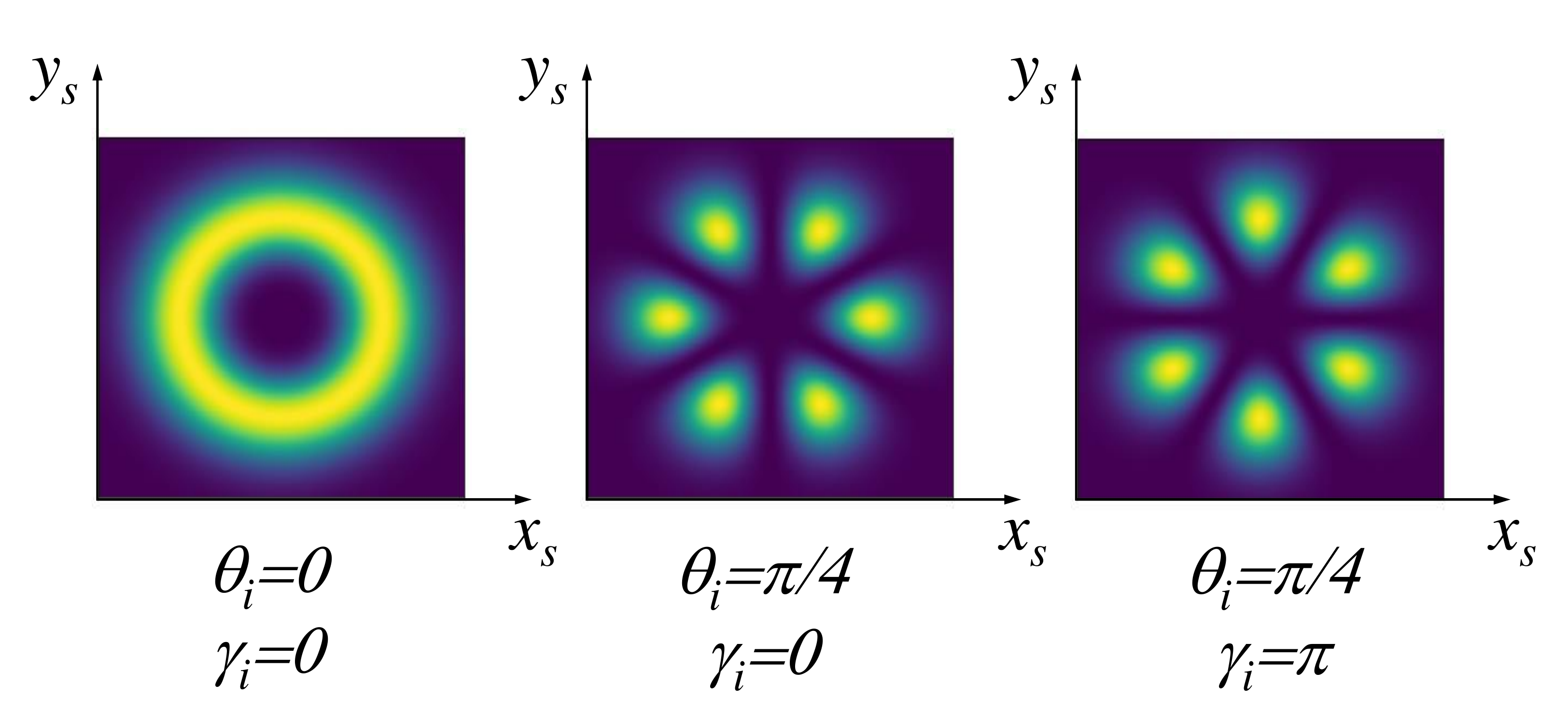}
	\caption{Laguerre-Gaussian coincidence pattern for an OAM pump beam with $l=3$ as a function of the 
		signal coordinates. The idler coordinates are fixed at the origin $x_i=y_i=0$ and the signal 
		polarization analyzer is set to $\gamma_s = 0$ and $\theta_s = \pi/4\,$.
		The LG coincidence pattern is modified as the idler polarization settings are changed.}
	\label{fig:LGimages}
\end{figure}

It is also interesting to see how orbital angular momentum affects the nonlocal correlations. For this end 
we assume the pump mode to be prepared in a superposition of Laguerre-Gaussian modes with zero radial 
order and OAM $-l$ with horizontal polarization and $+l$ with vertical polarization so that
\begin{eqnarray}
W_{V(H)}(\boldsymbol{\rho})=
\sqrt{\frac{2^{\abs{l}}}{\pi w^2}}
\left(\frac{x\pm iy}{w}\right)^{\!\abs{l}} 
e^{-(x^2+y^2)/w^2}\,.
\label{WLGmode}
\end{eqnarray}
Then, the coincidence counts are given by
\begin{eqnarray}
\!\!\!\!\!\!
&&C(\boldsymbol{\rho}_s,\boldsymbol{\rho}_i) =
\frac{K^2}{\pi\,w^4}\, e^{-\frac{2(x_s+x_i)^2}{w^2}} e^{-\frac{2(y_s+y_i)^2}{w^2}} \times
\label{CLG}\\
\!\!\!\!\!\!
&&
\left|\cos\theta_s \cos\theta_i\, \left[(x_s+x_i)+i(y_s+y_i)\right]^{\abs{l}} + \right.
\nonumber\\
\!\!\!\!\!\!
&&\left. \sin\theta_s\,\sin\theta_i\, e^{i(\gamma_s+\gamma_i)} \left[(x_s+x_i)-i(y_s+y_i)\right]^{\abs{l}}\right|^2 .
\nonumber
\end{eqnarray}
As before, the measurement strategy to  evidence the nonlocal spin-orbit correlations will be to fix the idler 
detector at its origin $x_i=y_i=0$ and to fix the signal polarization analyzer at $\theta_s=\pi/4$ and 
$\gamma_s=0\,$. Then, the coincidence pattern as a function of the signal coordinates becomes
\begin{eqnarray}
\!\!\!\!\!\!
&&C(\boldsymbol{\rho}_s,\boldsymbol{\rho}_i) =
\frac{K^2}{2\pi\,w^4}\, e^{-\frac{2(x_s^2+y_s^2)}{w^2}} \times
\label{CLG2}\\
\!\!\!\!\!\!
&&
\left|\cos\theta_i\, \left(x_s+iy_s\right)^{\abs{l}} + 
\sin\theta_i\, e^{i\gamma_i} \left(x_s-iy_s\right)^{\abs{l}}\right|^2 .
\nonumber
\end{eqnarray}
This result shows that any mode in the OAM Poincar\'e sphere \cite{Padgett1999} can be produced in 
the  coincidence pattern by scanning of the signal detector and varying  the idler polarization settings. A few 
examples are shown in Fig. \ref{fig:LGimages} for $l=3\,$, as measured in Ref.\cite{Samanta2017}. 
As the idler polarization settings are changed, the coincidence pattern is modified.

\subsection{OAM quantum correlations in two-photon vector vortex beams}
\label{sec:OAMcorr}

The two-photon spatial distribution can be written in the Laguerre-Gaussian basis\cite{spiral}. However, it is interesting to
extend this description for the case of the scheme with the two-photon-sandwich source. 
It will be useful for dealing with vector vortex pump beams. 
The LG modes are solutions of the paraxial wave equation in cylindrical coordinates. 
At the  beam waist plane ($z=0$), their mathematical expression in polar coordinates is
\begin{eqnarray}
&&\psi^{p}_{l}(\boldsymbol{\rho}) = R^{p}_{\abs{l}}(\rho)\, e^{il\phi}\,,
\nonumber\\
&&R^{p}_{\abs{l}}(\rho) = \sqrt{\frac{\left(2/\pi\right)\,p!}{\left(p+\abs{l}\right)!}}\,
\frac{\tilde{\rho}^{\abs{l}}}{w_0}\, 
L_p^{\abs{l}}(\tilde{\rho}^2)\, e^{-\tilde{\rho}^2/2}\,,
\nonumber\\
&&\tilde{\rho} = \sqrt{2}\,\rho/w_0\,,
\label{LGmodes}
\end{eqnarray}
where $p$ is the radial order, $l$ is the topological charge, $(\rho,\phi)$ are the transverse coordinates in the 
polar system, $w_0$ is the mode waist and $L^{\abs{l}}_p$ are the associated Laguerre polynomials. Pump, signal and 
idler beams are assumed to be mode matched, so that their wavefront radii are equal along the interaction length. 
This imposes a common Rayleigh distance $z_{Rp}=z_{Rs}=z_{Ri}\,$, which requires a different waist for each 
interacting beam according to $z_{R} = \pi w_{0j}^2/\lambda_j\,$.

The two-photon 
quantum state generated by the SPDC process can be cast as a superposition of different partitions of the 
pump OAM between signal and idler. First, let us derive the LG expansion of the correlated transverse 
momentum distribution of signal and idler, 
\begin{eqnarray}
v(\mathbf{q}_s + \mathbf{q}_i) = \sum_{\substack{p_1,l_1\\p_2,l_2}}\,A^{p_1 p_2}_{l_1 l_2}\,
\tilde{\psi}^{p_1}_{l_1}(\mathbf{q}_s)\,\tilde{\psi}^{p_2}_{l_2}(\mathbf{q}_i)\,,
\label{vexpansion}
\end{eqnarray}
where $\{\tilde{\psi}^{p}_{l}(\mathbf{q})\}$ are the LG mode functions in Fourier domain,
\begin{eqnarray}
\tilde{\psi}^{p}_{l}(\mathbf{q}) &=& \frac{1}{2\pi}
\int \psi^{p}_{l}(\boldsymbol{\rho}) \,e^{-i\mathbf{q}\cdot\boldsymbol{\rho}}\, d^2\boldsymbol{\rho}\;,
\nonumber\\
\psi^{p}_{l}(\boldsymbol{\rho}) &=& \frac{1}{2\pi}
\int \tilde{\psi}^{p}_{l}(\mathbf{q}) \,e^{i\mathbf{q}\cdot\boldsymbol{\rho}}\, d^2\mathbf{q}\;.
\end{eqnarray}
Both in the Fourier and position domains, the LG modes satisfy the following orthonormality 
\begin{eqnarray}
\int \psi^{p\,*}_{l}(\boldsymbol{\rho}) \,\psi^{p^{\prime}}_{l^{\prime}}(\boldsymbol{\rho})\, d^2\boldsymbol{\rho}
&=& \delta_{p p^{\prime}}\, \delta_{l l^{\prime}}\;,
\nonumber\\
\int \tilde{\psi}^{p\,*}_{l}(\mathbf{q}) \,\tilde{\psi}^{p^{\prime}}_{l^{\prime}}(\mathbf{q})\, d^2\mathbf{q}
&=& \delta_{p p^{\prime}}\, \delta_{l l^{\prime}}\;,
\end{eqnarray}
and completeness
\begin{eqnarray}
\sum_{p,l} \psi^{p\,*}_{l}(\boldsymbol{\rho}) \,\psi^{p}_{l}(\boldsymbol{\rho}^{\prime})
&=& \delta (\boldsymbol{\rho} - \boldsymbol{\rho}^{\prime})\;,
\nonumber\\
\sum_{p,l} \tilde{\psi}^{p\,*}_{l}(\mathbf{q}) \,\tilde{\psi}^{p}_{l}(\mathbf{q}^{\prime})
&=& \delta (\mathbf{q} - \mathbf{q}^{\prime})\;,
\end{eqnarray}
relations. By using them, the LG expansion coefficients in Eq. \eqref{vexpansion} are given by
\begin{eqnarray}
\!\!\!\!
A^{p_1 p_2}_{l_1 l_2} = \int v(\mathbf{q}_s + \mathbf{q}_i) \,
\tilde{\psi}^{p_1\,*}_{l_1}(\mathbf{q}_s)\,\tilde{\psi}^{p_2\,*}_{l_2}(\mathbf{q}_i)\,d^2\mathbf{q}_s \,d^2\mathbf{q}_i\;.
\label{coef-vexpansion}
\end{eqnarray}
These coefficients are more easily calculated in the position domain. This can be achieved by plugging the inverse 
Fourier transform of $\tilde{\psi}^{p}_{l}(\mathbf{q})$ and $v(\mathbf{q})$ into Eq. \eqref{coef-vexpansion} and 
using the Fourier representation of the Dirac delta function. The resulting expression is
\begin{eqnarray}
\!\!\!\!
A^{p_1 p_2}_{l_1 l_2} = \int W(\boldsymbol{\rho}) \,
\psi^{p_1\,*}_{l_1}(\boldsymbol{\rho})\,\psi^{p_2\,*}_{l_2}(\boldsymbol{\rho})\,d^2\boldsymbol{\rho}\,.
\label{coef-vexpansion-overlap}
\end{eqnarray}
With the expansion coefficients in hands, we can rewrite the two-photon state \eqref{phitau} in the Fock 
basis of OAM modes
\begin{eqnarray}
\ket{\Phi}_{si} = \kappa \sum_{\substack{p_1,l_1\\p_2,l_2}} 
\left[
A^{p_1 p_2}_{l_1 l_2} \ket{p_1,l_1,H}_s \ket{p_2,l_2,H}_i 
\right.
\nonumber\\
\left. + \,\,B^{p_1 p_2}_{l_1 l_2} \ket{p_1,l_1,V}_s \ket{p_2,l_2,V}_i \right] 
\!,
\label{phitauOAM}
\end{eqnarray}
where 
\begin{eqnarray}
\!\!\!\!\!\!\!\!\!\!\!\!
&&\ket{p,l,\mu} = \int d^2\mathbf{q}\,\, \tilde{\psi}^{p}_{l}(\mathbf{q}) \,\ket{1_{\mathbf{q},\mu}}
\qquad (\mu=H,V)\,,
\nonumber\\
&&\braket{p,l,\mu|p^{\prime},l^{\prime},\mu^{\prime}}=\delta_{p p^{\prime}}\delta_{l l^{\prime}}\delta_{\mu \mu^{\prime}}\,,
\label{OAM-Fock}
\end{eqnarray}
are single-photon OAM states with polarization $\mu$ and the coefficients 
$A^{p_1 p_2}_{l_1 l_2}$ and $B^{p_1 p_2}_{l_1 l_2}$ are given by 
Eq. \eqref{coef-vexpansion-overlap} with $W_V$ and $W_H\,$, respectively.

\begin{figure}
	\includegraphics[scale=0.4]{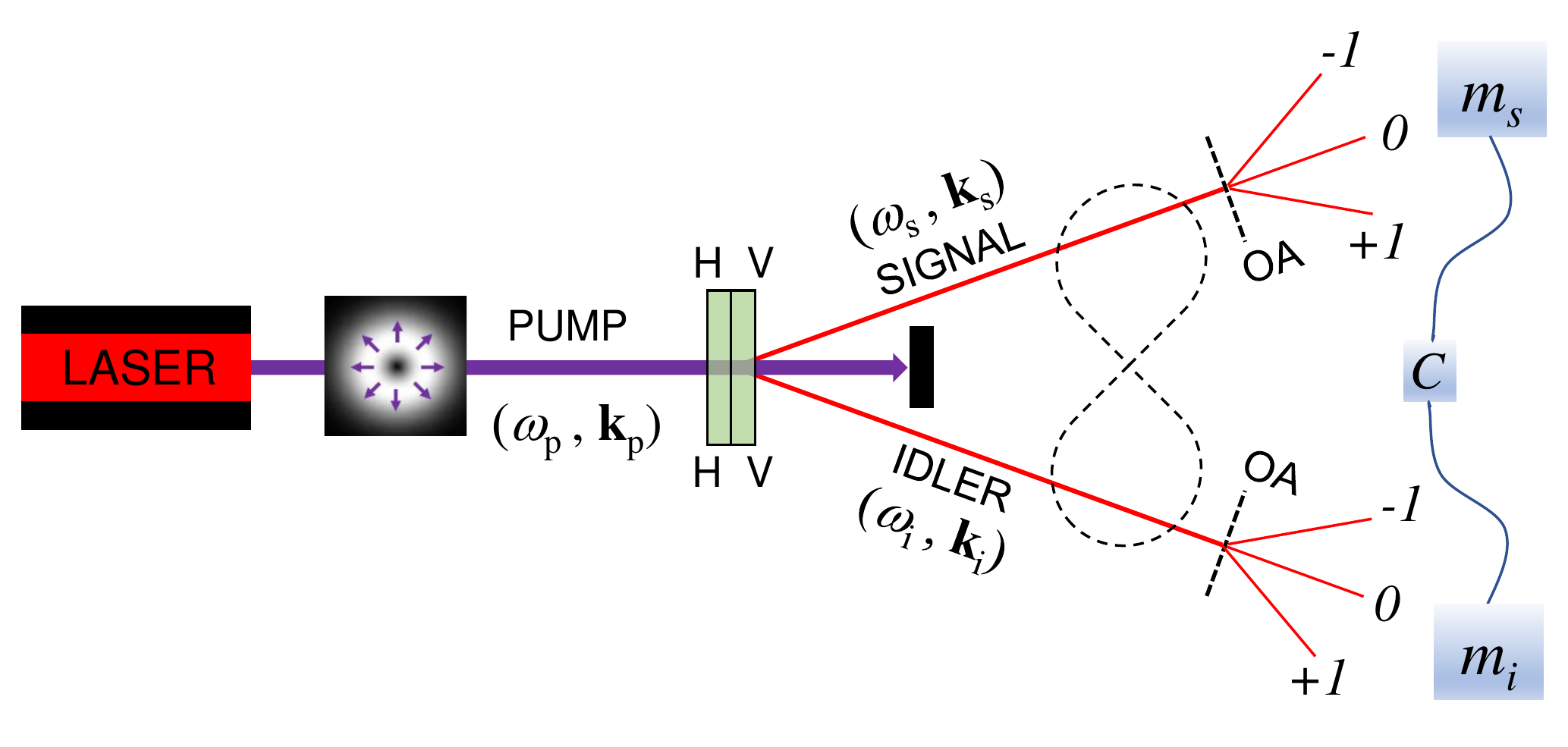}
	\caption{Setup for OAM correlation measurements. An OAM analyzer (OA) is used in each detection arm, 
		allowing for correlation measurements between different outputs of signal and idler, with topological 
		charges constrained by $m_{s} + m_{i} = \pm l\,$.}
	\label{fig:setupOAM}
\end{figure}

Let us assume that the pump 
beam is prepared in a vector mode of the kind considered in the second example of section 
\ref{sec:nonlocalvb}, a superposition of the LG mode $(0,-l)$ with horizontal 
polarization and mode $(0,+l)$ with vertical polarization 
\begin{eqnarray}
\boldsymbol{\Psi}(\boldsymbol{\rho}) = \frac{\psi^0_{-l}(\boldsymbol{\rho})\,\mathbf{\hat{e}}_H + 
	\psi^0_{+l}(\boldsymbol{\rho})\,\mathbf{\hat{e}}_V}{\sqrt{2}}\;.
\end{eqnarray}
In this case, we have set $W_V=\psi^0_{+l}/\sqrt{2}$ and $W_H=\psi^0_{-l}/\sqrt{2}$ and the OAM expansion 
coefficients are 
\begin{eqnarray}
\!\!\!\!\!\!\!\!\!\!\!\!\!\!\!\!\!\!\!\!
&&A^{p_1 p_2}_{l_1 l_2} = \sqrt{2}\pi \delta_{l,l_1+l_2} \!\!\int\!\! R^{0}_{\abs{l}}(\rho) \,
R^{p_1\,*}_{\abs{l_1}}(\rho)\,R^{p_2\,*}_{\abs{l-l_1}}(\rho)\,\rho\, d\rho ,
\nonumber\\
\!\!\!\!\!\!\!\!\!\!\!\!\!\!\!\!\!\!\!\!
&&B^{p_1 p_2}_{l_1 l_2} = \sqrt{2}\pi \delta_{-l,l_1+l_2} \!\!\int\!\! R^{0}_{\abs{l}}(\rho) \,
R^{p_1\,*}_{\abs{l_1}}(\rho)\,R^{p_2\,*}_{\abs{l+l_1}}(\rho)\,\rho\, d\rho ,
\label{coef-vexpansion-OAM}
\end{eqnarray}
where the Kronecker deltas, $\delta_{l,l_1+l_2}$ and $\delta_{-l,l_1+l_2}\,$, result from the angular integration. 
They impose the OAM conservation condition in the two-photon state \eqref{phitauOAM}. Moreover, the following 
symmetry relations hold
\begin{eqnarray}
A^{p_1 p_2}_{m \,l-m} &=& B^{p_1 p_2}_{-m \,m-l}\;,
\nonumber\\
A^{p_1 p_2}_{m \,l-m} &=& A^{p_2 p_1}_{l-m \,m}\;.
\end{eqnarray}
They allow us to rewrite the two-photon state in a 
more convenient way that makes more evident the simultaneous OAM and polarization entanglement
\begin{eqnarray}
\!\!\!\!\!\!\!
&&\ket{\Phi}_{si} = \kappa \!\sum_{p_1,p_2,m}\!
A^{p_1 p_2}_{m \,l-m} \times
\label{phitauOAM2}\\
\!\!\!\!\!\!\!
&&\left(\frac{}{}\!\!\ket{p_1,m,H}\ket{p_2,l-m,H} + \ket{p_1,-m,V}\ket{p_2,m-l,V}\right).
\nonumber
\end{eqnarray}
This form of the two-photon vector beam quantum state exhibits explicitly the simultaneous OAM 
and polarization entanglement. It is useful for measurement schemes where OAM  sorting is 
implemented in each detection arm, as depicted in Fig. \ref{fig:setupOAM}.  While this representation
of the OAM sorter is idealized for pedagogic purposes, there are several types of architectures being
developed to this end, meaning that the OAM sorting opeartion is already viable with increasing
efficiency and resolution\cite{Kishikawa_2018}.

\begin{figure}
	\includegraphics[scale=0.6]{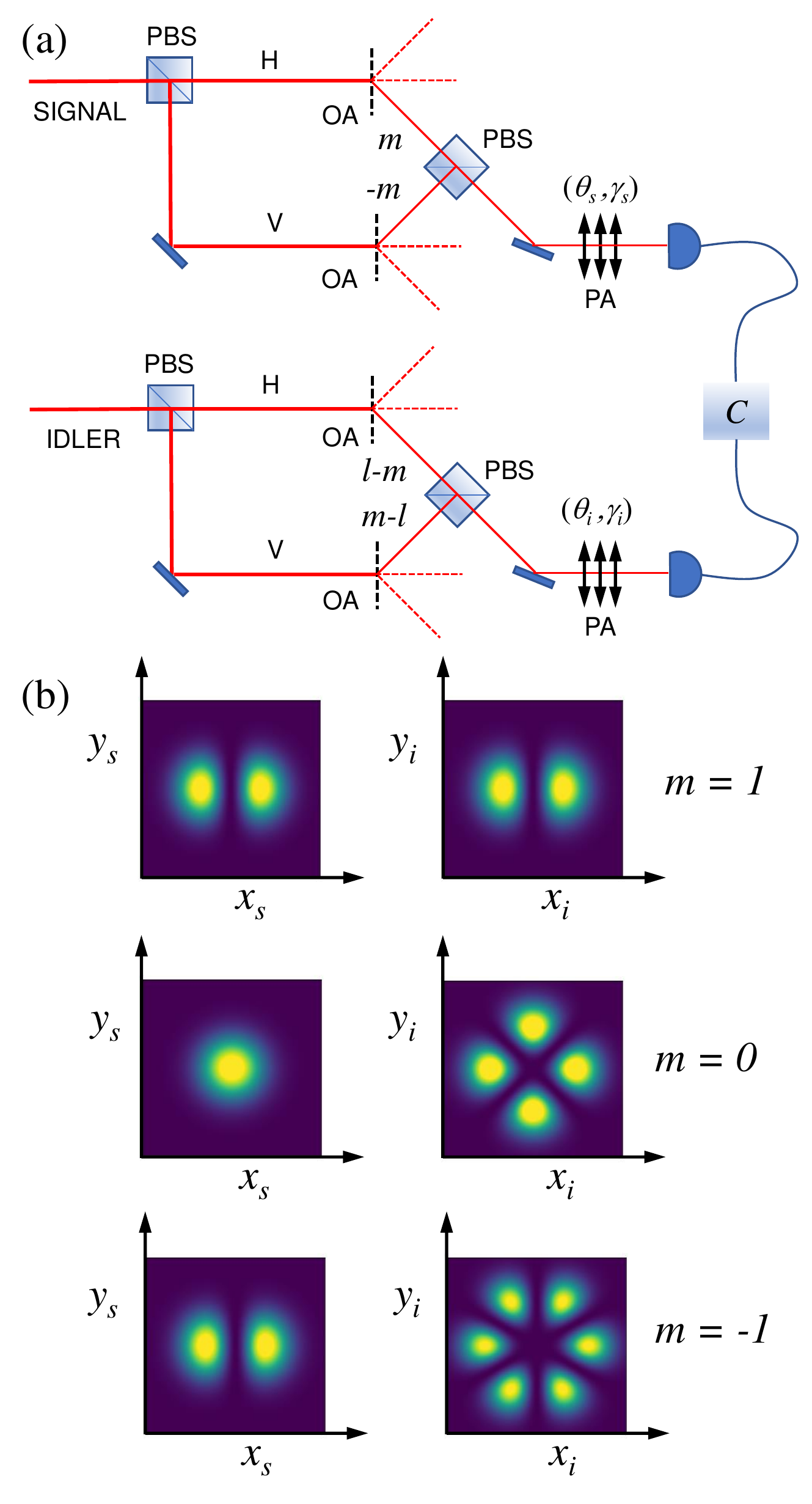}
	\caption{(a) Setup for measuring the OAM correlated images $|\mathcal{F}^{0 0}_{m\,l-m} (\boldsymbol{\rho}_s,\boldsymbol{\rho}_i)|^2$. 
		PBS: polarizing beam splitter, PA: polarization analyzer, OA: OAM analyzer.
		(b) Correlation images between different OAM outputs of signal and idler with topological charges constrained 
		by $m_{s} + m_{i} = \pm l$ and $p_1=p_2=0\,$. In this example we have set the pump topological charge $l=2$ and 
		polarization analysis is set to $\theta_s=\theta_i=\pi/4$ and $\gamma_s=\gamma_i=0\,$. The images at the signal (idler) 
		plane are calculated with the idler (signal) detector fixed at $x_{i}=w_0\,\sqrt{\abs{l-m}/2}\,$, $y_{i}=0$ 
		($x_{s}=w_0\,\sqrt{\abs{m}/2}\,$, $y_{s}=0$) for optimal coincidence amplitude.}
	\label{fig:setup-LGimages}
\end{figure}

It is instructive to obtain the OAM decomposition of the spatial correlations. We start by writing the 
positive frequency component of the electric field operator in terms of annihilation operators 
$a^j_{pl}$ and $b^j_{pl}$ ($j=s,i$) of photons in Laguerre-Gaussian modes $(p,l)$ with horizontal and 
vertical polarizations, respectively, 
\begin{eqnarray}
\hat{E}^{+}_{jH} &=& \sum_{pl} a^{j}_{pl}\,\psi^p_l(\boldsymbol{\rho})\,,
\nonumber\\
\hat{E}^{+}_{jV} &=& \sum_{pl} b^{j}_{pl}\,\psi^p_l(\boldsymbol{\rho})\,.
\end{eqnarray}
These annihilation operators are related to those in the transverse momentum basis through 
\begin{eqnarray}
a^j_H (\mathbf{q}) &=& \frac{1}{2\pi i\mathcal{E}_j}\sum_{pl} a^{j}_{pl}\,\tilde{\psi}^p_l(\mathbf{q})\,,
\nonumber\\
a^j_{pl} &=& 2\pi i\mathcal{E}_j \int a^j_H (\mathbf{q}) \,\tilde{\psi}^{p\,*}_l(\mathbf{q})\, d^2\mathbf{q} \,,
\end{eqnarray}
and the equivalent relations for $a^j_V (\mathbf{q})$ and $b^j_{pl}\,$. The action of the OAM annihilation operators 
on the corresponding Fock states is given by
\begin{eqnarray}
a^j_{pl}\ket{p^{\prime},l^{\prime},H} &=& 
b^j_{pl}\ket{p^{\prime},l^{\prime},V} =
\delta_{p p^{\prime}}\delta_{l l^{\prime}} \ket{\mathrm{vac}} \,,
\nonumber\\
a^j_{pl}\ket{p^{\prime},l^{\prime},V} &=& 
b^j_{pl}\ket{p^{\prime},l^{\prime},H} = 0\,.
\end{eqnarray}
In terms of the OAM eigenfunctions, the auxiliary vectors defined in Eq. \eqref{betamu} for calculating the 
quantum correlations become
\begin{eqnarray}
\ket{\beta_{H}} &=& \hat{E}^+_{iH}(\boldsymbol{\rho}_i)\hat{E}^+_{sH}(\boldsymbol{\rho}_s)\ket{\Phi}_{si}
\nonumber\\
&=& \kappa \!\sum_{p_1,p_2,m}\! A^{p_1 p_2}_{m\,l-m}\, 
\psi^{p_1}_{m}(\boldsymbol{\rho}_s) \psi^{p_2}_{l-m}(\boldsymbol{\rho}_i)
\ket{\mathrm{vac}}\,,
\nonumber\\
\ket{\beta_{V}} &=& \hat{E}^+_{iV}(\boldsymbol{\rho}_i)\hat{E}^+_{sV}(\boldsymbol{\rho}_s)\ket{\Phi}_{si}
\label{betamuOAM}\\
&=& \kappa \!\sum_{p_1,p_2,m}\! A^{p_1 p_2}_{m\,l-m}\, 
\psi^{p_1}_{-m}(\boldsymbol{\rho}_s) \psi^{p_2}_{m-l}(\boldsymbol{\rho}_i)
\ket{\mathrm{vac}}\,.
\nonumber
\end{eqnarray}
Finally, the coincidence counts can be calculated from Eq. \eqref{cbeta} giving 
\begin{eqnarray}
\!\!\!\!\!\!\!\!\!\!\!\!\!\!\!\!
&&C(\boldsymbol{\rho}_s,\boldsymbol{\rho}_i) = \kappa^2\,
\left| \sum_{p1,p2,m} A^{p_1 p_2}_{m\,l-m} \mathcal{F}^{p_1 p_2}_{m\,l-m} (\boldsymbol{\rho}_s,\boldsymbol{\rho}_i) \right|^2\,,
\label{C-OAM2}
\end{eqnarray}
where
\begin{eqnarray}
\!\!\!\!\!\!\!\!
\mathcal{F}^{p_1 p_2}_{m\,l-m} (\boldsymbol{\rho}_s,\boldsymbol{\rho}_i) \!&=&\!
\cos\theta_s \cos\theta_i\, \psi^{p_1}_{m}(\boldsymbol{\rho}_s) \psi^{p_2}_{l-m}(\boldsymbol{\rho}_i) + 
\\
\!\!\!\!\!\!\!\!
&+&\! e^{i\gamma_+}\, \sin\theta_s \sin\theta_i\, 
\psi^{p_1}_{-m}(\boldsymbol{\rho}_s) \psi^{p_2}_{m-l}(\boldsymbol{\rho}_i)\,,
\nonumber
\end{eqnarray}
and $\gamma_+\equiv \gamma_s + \gamma_i\,$.
Note that Eq. \eqref{C-OAM2} is the Schmidt decomposition of the coincidence pattern derived in Eq. \eqref{CLG} in 
terms of factorized OAM eigenfunctions for signal and idler. The different components 
$\mathcal{F}^{p_1 p_2}_{m\,l-m} (\boldsymbol{\rho}_s,\boldsymbol{\rho}_i)$
can be accessed by mode sorting before each detector, as indicated in Fig. \ref{fig:setup-LGimages}a. 
First, the $H$ and $V$ polarizations of signal and idler are separated by polarizing beam splitters (PBS) 
and pass through OAM sorters. Then, the $m$ component with $H$ polarization is recombined with the 
$-m$ component with $V$ polarization in a second PBS at the signal arm. In the same way, the $l-m$ component
with $H$ polarization and $m-l$ with $V$ polarization are recombined at the idler arm. 
Finally, the polarization information is erased in each arm by analyzers (PA). 

Note that two correlation channels are involved. The OAM components of $H-$polarized photons add 
up to $l$ while those of $V-$polarized photons add up to $-l\,$. After polarization erasure, 
the two channels interfere and the resulting coincidence pattern is 
$|\mathcal{F}^{p_1 p_2}_{m\,l-m} (\boldsymbol{\rho}_s,\boldsymbol{\rho}_i)|^2$. 
They differ 
fundamentally from those calculated in section \ref{sec:nonlocalvb} (see Fig. \ref{fig:LGimages}). There, 
similar patterns are obtained when either $\boldsymbol{\rho}_s$ or $\boldsymbol{\rho}_i$ is scanned while the 
other remains fixed. Here, due to transverse mode analysis before detection, the resulting coincidence patterns 
will present different shapes, depending on which detector is scanned. In Fig. \ref{fig:setup-LGimages}b 
we show different OAM correlated images $|\mathcal{F}^{0 0}_{m\,l-m} (\boldsymbol{\rho}_s,\boldsymbol{\rho}_i)|^2$ 
for a pump topological charge $l=2$ and polarization analysis set to $\theta_s=\theta_i=\pi/4\,$, 
$\gamma_s=\gamma_i=0\,$. 

\section{conclusion}
\label{sec:conclusion}

In conclusion, we present the quantum theory of the angular spectrum transfer in spontaneous parametric down-conversion, generalized for structured light beams. The scheme studied is based on a source composed by two non-linear crystals to produce simultaneous nonlocal 
correlations in the spin and orbital angular momentum of entangled photon pairs. A structured pump beam, carrying internal non-separability between its polarization and transverse mode, generates photon pairs that are quantum correlated in both polarization and transverse spatial degrees of freedom. The two-photon spin-orbit quantum state generated by the process is derived and the corresponding quantum correlations are calculated both in position and OAM domains. The results were obtained in the weak interaction regime that allows us to safely neglect higher order terms, keeping only two-photon events. In cases where the interaction strength increases, the presence of four-photon events may become significant, leading to the reduction of the purity of the two-photon states and reduction of the entanglement in both polarization and transverse spatial degrees of freedom.

We show how the spatial correlations between signal and idler fields can be shaped by different settings of remote polarizers placed before their respective detectors. As an example, we show that Hermite-Gaussian and Laguerre-Gaussian coincidence distributions over the signal coordinates can be transformed by the remote control of the idler polarization settings. This resembles the behavior of a vector beam and, in this sense, we interpret this type of two-photon spin-orbit structure as a non-local vector beam. One natural follow up of the present work is the application of this structure to alignment-free quantum cryptography with non-local correlations \cite{Souza2008,DAmbrosio2012,Ekert1991}. Another promising application concerns the use of machine-learning-based protocols to classify non local vector vortex beams. One potential route to this application is the combination of the classification scheme for vector vortex beams introduced in Ref. \cite{Giordani2020} with the single photon wave front correction scheme demonstrated in Ref. \cite{Bhusal2020}, where a triggering photon heralds the single photon populating a Laguerre Gaussian mode. In the extension of the approach to non local vector vortex beams, polarization measurements in the triggering photon would allow the classification of the two-photon state. This is a natural resource for quantum communication systems.
Moreover, the Schmidt decomposition in terms of Laguerre-Gaussian modes gives rise to different types of polarization dependent spatial correlations that can be accessed with mode filtering techniques.

\section*{Acknowledgments}

The authors would like to thank the Brazilian Agencies, Conselho Nacional de Desenvolvimento Tecnol\'ogico (CNPq), Funda\c c\~{a}o Carlos Chagas Filho de Amparo \`{a} 
Pesquisa do Estado do Rio de Janeiro (FAPERJ), Funda\c c\~{a}o de Amparo \`{a} 
Pesquisa do Estado de Santa Catarina (FAPESC) and the Brazilian National Institute of Science and Technology of Quantum Information (INCT/IQ). This study was financed in part by the Coordena\c c\~{a}o de Aperfei\c coamento de Pessoal de N\'ivel Superior - Brasil (CAPES) - Finance Code 001.

\newpage
\appendix

\section{Individual Intensities}
\label{individual}

We will workout explicitly the intensity of horizontally polarized signal photons. The extension to vertical polarization 
and to idler photons is straightforward. 
\begin{eqnarray}
I_j &=& \norm{\cos\theta_j \ket{\alpha_{jH}} + \sin\theta_j \,e^{i\gamma_j}\ket{\alpha_{jV}}}^2\;,
\end{eqnarray}

The corresponding auxiliary vector becomes 
\begin{eqnarray}
\ket{\alpha_{sH}} = i \kappa\mathcal{E}_{s} \int d^2\mathbf{q}_i\, 
G_{V}(\mathbf{q}_i,\boldsymbol{\rho}_s) \ket{1_{\mathbf{q}_{i},V}}\,,
\end{eqnarray}
where
\begin{eqnarray}
G_{V}(\mathbf{q}_i,\boldsymbol{\rho}_s) &\equiv& \int d^2\mathbf{q}_s\, v^p_{V}(\mathbf{q}_s + \mathbf{q}_i) \,
e^{i\left[\mathbf{q}_s\cdot\boldsymbol{\rho}_s + \varphi(\mathbf{q}_s)\right]}\qquad
\\
&=& e^{-i\left(\mathbf{q_i}\cdot\boldsymbol\rho_s+\frac{q_i^2}{2k_s}z_s\right)}\,
W_{V}\left(\boldsymbol{\rho}_s + \frac{z_s}{2k_s}\mathbf{q}_i  \,,z_s\right)\,.
\nonumber
\end{eqnarray}
Using the orthonormality condition for the Fock states 
$\braket{n_{\mathbf{q},\mu}|m_{\mathbf{q}^{\prime},\nu}} = 
\delta_{nm}\,\delta_{\mu\nu}\,\delta(\mathbf{q}-\mathbf{q}^{\prime})\,$,
we have $\braket{\alpha_{sH}|\alpha_{sV}}=0$ and
\begin{eqnarray}
\braket{\alpha_{sH}|\alpha_{sH}} &=& \kappa^2\mathcal{E}_{s}^2 \,
\int d^2\mathbf{q}_i
\abs{W_{V}\left(\boldsymbol{\rho}_s + \frac{z_s}{2k_s}\mathbf{q}_i  \,,z_s\right)}^2 
\nonumber\\
&=& \kappa^2\mathcal{E}_{s}^2
\int d^2\boldsymbol{\rho}\, \abs{W_{V}\left(\boldsymbol{\rho},0\right)}^2 \,,
\end{eqnarray}
assuming $W_{V}$ to be normalizable. 
Along the same lines we can easily arrive at
\begin{eqnarray}
\braket{\alpha_{sV}|\alpha_{sV}} &=& \kappa^2\mathcal{E}_{s}^2 \,
\int d^2\mathbf{q}_i
\abs{W_{H}\left(\boldsymbol{\rho}_s + \frac{z_s}{2k_s}\mathbf{q}_i  \,,z_s\right)}^2 
\nonumber\\
&=& \kappa^2\mathcal{E}_{s}^2
\int d^2\boldsymbol{\rho}\, \abs{W_{H}\left(\boldsymbol{\rho},0\right)}^2 \,.
\end{eqnarray}
Moreover,the same deduction can be applied to the idler individual intensity and we finally get
\begin{equation}
I_j = \cos^2\theta_j \mathcal{I}_{H} + \sin^2\theta_j \mathcal{I}_{V}\,,
\label{app-individual}
\end{equation}
with $j=s,i$ and  
\begin{equation}
\mathcal{I}_{H(V)} = \kappa^2\mathcal{E}_{s}^2 \, \int d^2\boldsymbol{\rho}
\abs{W_{V(H)}\left(\boldsymbol{\rho}\right)}^2 \,.
\end{equation}
Note that after integration, no spatial dependence is left in the individual intensities. This 
means that the pump spatial properties are washed out in the individual intensity of the signal beam. 

\section{Coincidence Counts}
\label{coincidence}

The coincidence count can be obtained from the following norm, 

\begin{eqnarray}
\!\!\!\!\!\!\!\!\!\!\!\!\!\!\!\!\!\!
&&C(\boldsymbol{\rho}_s,\boldsymbol{\rho}_i) =
\label{cbeta}\\
\!\!\!\!\!\!\!\!\!\!\!\!\!\!\!\!\!\!
&&\left|\left|\!\!\frac{}{}\cos\theta_s\cos\theta_i \ket{\beta_{H}} \right.\right.
+ \left.\left. \sin\theta_s\sin\theta_i \,e^{i(\gamma_s + \gamma_i)} \ket{\beta_{V}}\right|\right|^2 .
\nonumber
\end{eqnarray}
This norm can be calculated by writing the auxiliary vectors in terms of the vacuum state 
as follows,
\begin{eqnarray}
\!\!\!\!\!\!\!\!\!\!\!\!\!
\ket{\beta_{\mu}} &=& - \kappa\mathcal{E}_{s}\mathcal{E}_{i} \int d^2\mathbf{q}_s \int d^2\mathbf{q}_i\, 
v^p_{\mu^{\prime}}(\mathbf{q}_s + \mathbf{q}_i)
\nonumber\\
&\times& \,e^{i\left[\mathbf{q}_s\cdot\boldsymbol{\rho}_s + \mathbf{q}_i\cdot\boldsymbol{\rho}_i + 
	\varphi(\mathbf{q}_s) + \varphi(\mathbf{q}_i)\right]} \,\ket{\mathrm{vac}}
\nonumber\\
&=&- \kappa\mathcal{E}_{s}\mathcal{E}_{i} \abs{J}^2\int d^2\mathbf{q}_- \, 
e^{i\left[\mathbf{q}_-\cdot\boldsymbol{\rho}_- + \varphi(\mathbf{q}_-)\right]}
\nonumber\\
&\times&\int d^2\mathbf{q}_+\, 
v^p_{\mu^{\prime}}(\mathbf{q}_+) \,
e^{i\left[\mathbf{q}_+\cdot\boldsymbol{\rho}_+ + \varphi(\mathbf{q}_+)\right]}\,\ket{\mathrm{vac}}\,,
\label{betamu+-}
\end{eqnarray}
where $\mu\neq\mu^{\prime}$ and we defined 
\begin{eqnarray}
&& \mathbf{q}_{+} = \mathbf{q}_s + \mathbf{q}_i \,,
\qquad\qquad\;\;
\mathbf{q}_{-} = \frac{k_i z_s}{k_s z_i} \,\mathbf{q}_s - \mathbf{q}_i \,,
\nonumber\\
&& \boldsymbol{\rho}_{+} = \frac{k_s z_i \,\boldsymbol{\rho}_s + k_i z_s \,\boldsymbol{\rho}_i}{k_s z_i + k_i z_s}\,,
\;\;
\boldsymbol{\rho}_{-} = \frac{k_i z_s \,\left(\boldsymbol{\rho}_s - \,\boldsymbol{\rho}_i\right)}{k_s z_i + k_i z_s}\,,\qquad
\nonumber\\
&& \varphi(\mathbf{q}_+) = k_s z_s + k_i z_i - \left(\frac{k_s z_i}{k_s z_i + k_i z_s}\right)\,\frac{z_s}{2k_s}\, q_+^2\,,
\nonumber\\
&& \varphi(\mathbf{q}_-) = -\left(\frac{k_s z_i}{k_s z_i + k_i z_s}\right)\,\frac{z_i}{2k_i}\, q_-^2\,,
\nonumber\\
&& J = -\frac{k_s z_i}{k_s z_i + k_i z_s}\,.
\end{eqnarray}
The integrals in \eqref{betamu+-} are considerably simplified by assuming $z_s = z_i = z$ and using 
$k_s + k_i = k_p\,$,
\begin{eqnarray}
&& \int d^2\mathbf{q}_- \, 
e^{i\left[\mathbf{q}_-\cdot\boldsymbol{\rho}_- + \varphi(\mathbf{q}_-)\right]}
= e^{-i \frac{k_p k_i}{k_s}\frac{\rho_-^2}{2z}}\,,
\nonumber\\
&& \int d^2\mathbf{q}_+\, 
v^p_{\mu^{\prime}}(\mathbf{q}_+) \,
e^{i\left[\mathbf{q}_+\cdot\boldsymbol{\rho}_+ + \varphi(\mathbf{q}_+)\right]} 
= W_{\mu^{\prime}} \left(\boldsymbol{\rho}_+,z\right)\,. \qquad
\label{integrals+-}
\end{eqnarray}
The integration on $\mathbf{q}_-$ gives an irrelevant phase factor, while that on $\mathbf{q}_+$ 
brings the pump spatial properties as a function of the joint positions of signal and idler.
Thus, we arrive at the following result for the auxiliary vectors 
\begin{eqnarray}
&&\ket{\beta_{\mu}} =
- \kappa\mathcal{E}_{s}\mathcal{E}_{i} \abs{J}^2\,e^{-i \frac{k_p k_i}{k_s}\frac{\rho_-^2}{2z}}\,
W_{\mu^{\prime}} \left(\boldsymbol{\rho}_+,z\right)\,\ket{\mathrm{vac}}\,.\qquad
\label{betamuW}
\end{eqnarray}
Finally, the intensity correlations are given by
\begin{eqnarray}
\!\!\!\!\!\!\!
C(\boldsymbol{\rho}_s,\boldsymbol{\rho}_i) &=& K^2
\left|\!\!\frac{}{}\cos\theta_s\cos\theta_i W_{V}\left(\boldsymbol{\rho}_+,z\right) \right.
\nonumber\\
\!\!\!\!\!\!\!
&+& \left. 
\sin\theta_s\sin\theta_i \,e^{i(\gamma_s + \gamma_i)} W_{H}\left(\boldsymbol{\rho}_+,z\right)\right|^2 .
\end{eqnarray}
where $K \equiv \kappa\mathcal{E}_{s}\mathcal{E}_{i} \abs{J}^2\,$. 

\bibliography{q-vector-beams-Bib}

\end{document}